\documentclass[options]{JHEP3}
\usepackage{amsmath}

\title{Mixed heavy-quark--gluon condensate in the stochastic vacuum model and dual superconductor}

\author{Dmitri Antonov\thanks{Permanent address: 
        ITEP, B. Cheremushkinskaya 25, RU-117 218 Moscow, Russia.}\\ 
        Institute of Physics, Humboldt University of Berlin\\
        Newtonstr. 15, 12489 Berlin, Germany \\       
        E-mail: \email{antonov@physik.hu-berlin.de}}

\abstract{The world-line formalism is used for the evaluation of 
the mixed heavy-quark--gluon condensate in two models of QCD -- the stochastic vacuum model and the dual superconductor one.
Calculations are performed for an arbitrary dimensionality of space-time $d\ge 2$.
While in the stochastic vacuum model, 
the condensate is UV finite up to $d=8$, in the dual superconductor model
it is UV divergent at any $d\ge 2$. 
A regularization of this divergence is proposed, which makes quantitative the condition of the type-II dual superconductor.
The obtained results are generalized to the case of finite temperatures. Corrections to the both, mixed and standard, heavy-quark 
condensates, which appear due to the variation of the gauge field at the scale of the vacuum correlation length, are evaluated within the 
stochastic vacuum model. These corrections diminish the absolute values of the condensates, as well as the ratio of the mixed condensate to the 
standard one.}

\keywords{nonperturbative effects; confinement; QCD; phenomenological models}

\preprint{HU-EP-05/32}

\dedicated{}
\begin{document}

\section{Introduction and the idea of the method}
The mixed quark-gluon condensate, $g\left<\bar\psi\sigma_{\mu\nu}F_{\mu\nu}\psi\right>$, which by its definition measures the average interaction 
of the color-magnetic moment of a quark with the vacuum fields, plays an important role in QCD sum rules~\cite{5} (for a review see~\cite{6}).
Lattice measurements of this quantity in the chiral limit have been performed~\cite{7}. 
Surprisingly, an {\it analytic} formula for this condensate has been derived only quite recently~\cite{8}, 
within the stochastic vacuum model of QCD~\cite{90, ds} (for a recent review see~\cite{110}).
In this paper, we will use the method of ref.~\cite{4} to evaluate the mixed condensate in the 
heavy-quark limit (i.e. for $b$, $c$, and $t$ quarks), and for an arbitrary number of space-time dimensions. The method is based on the 
so-called world-line formalism~\cite{1}, 
which is by now a well developed tool for {\it perturbative} calculations in field theory (for a review see~\cite{2}).
Besides that, various attempts exist to apply the method to {\it nonperturbative} problems as well~\cite{3,4}. Here, we will thus 
continue this line of research.

The idea, which makes the world-line calculations possible analytically, is to transform the world-line integral
to the one in an effective {\it constant} Abelian field, to be 
averaged over. The weight of the average, being prescribed by the form of the heavy-quark 
Wilson loop in the {\it original} confining theory, inherits therefore the information about that theory.
The area of the minimal
surface bounded by the quark trajectory ${\cal C}$, $S_{\rm min}({\cal C})$, is $\lesssim m^{-2}$, where $m$ is the current quark mass.
For heavy quarks, $S_{\rm min}({\cal C})$ becomes 
smaller than the squared  
vacuum correlation length, $T_g^2$, since $T_g\sim 1{\,}{\rm GeV}^{-1}$~\cite{11, 12}. 
(This length is defined as a
distance at which the two-point gauge-invariant correlation function
of $F_{\mu\nu}$'s falls off at least exponentially.)
For such small-sized Wilson loops, one can write 
\begin{equation}
\label{appr}
S_{\rm min}^2\simeq\frac12\Sigma_{\mu\nu}^2.
\end{equation} 
Here, $\Sigma_{\mu\nu}=\oint_{\cal C}^{}x_\mu dx_\nu$ is
the so-called tensor area, and 
``$\simeq$'' means ``for nearly flat contours'', since eq.~(\ref{appr}) is exact only when ${\cal C}$ is flat and lies in the $(\mu\nu)$-plane.

In the known confining theories, the following formulae for a small-sized
Wilson loop exist~\footnote{As in~\cite{8,4}, $\left<W({\cal C})\right>$, which we consider in this paper, 
is a purely nonperturbative part of the 
full Wilson loop, i.e. the part, which can generate the terms starting from the linear one in the quark-antiquark potential.
The Coulombic term corresponds to the one-gluon exchange, which yields the multiplicative renormalization of the loop. The corresponding 
renormalization factor is referred to 
the definition of the path-integral measure in eq.~(\ref{1}) below.}: 

$\bullet$ Stochastic vacuum model~\cite{ds}~\footnote{This expression will be proven at the beginning of subsection~2.1.}:
\begin{equation}
\label{quadr}
\left<W({\cal C})\right>\simeq N\exp\left(-C\Sigma_{\mu\nu}^2\right),
\end{equation}
where $C\equiv\frac{g^2\left<F^2\right>}{8N(d^2-d)}$, 
$\left<F^2\right>\equiv\left<(F_{\mu\nu}^a(0))^2\right>$;

$\bullet$ Abelian-type theories with confinement:
\begin{equation}
\label{lin}
\left<W({\cal C})\right>\simeq N\exp\left(-\sigma\sqrt{\frac12
\Sigma_{\mu\nu}^2}\right).
\end{equation}
To the latter theories belong the 3d weakly coupled Georgi-Glashow model~\cite{polyakov} and the 4d dual Abelian Higgs model.
In this paper, we will consider a SU($N$)-generalization of the latter model, which was proposed for $N=3$ in ref.~\cite{ms} and generalized 
to arbitrary $N$ in ref.~\cite{a}. This is just a dual superconductor model of 
confinement, based on the monopole condensation. With the logarithmic accuracy, the string tension of a short string in this theory reads
$\sigma=2\pi(N-1)\eta^2\ln\kappa$, where $\eta$ is the vacuum expectation
value of the dual Higgs field, and $\kappa$ is the Landau-Ginzburg parameter; $\ln\kappa\gg 1$ in the London limit under study.

Note that, in QCD at small distances, it is unlikely to have such a form of the Wilson loop. 
Indeed, under the assumption that the Feynman-Kac formula, which relates the
potential of a heavy $q\bar q$-pair to the Wilson loop, can be
extrapolated down to the distances smaller than $T_g$, this would correspond to the linear
next-to-Coulombic term in the potential. Such a term is however ruled out in QCD, as indicated by the 
pNRQCD calculations~\cite{pNR}, the world-line calculations of the associated tachyonic gluon mass~\cite{tach}, as well as the lattice data~\cite{sommer}.
As a support of this conclusion, it has been shown in ref.~\cite{4} that the heavy-quark condensate $\left<\bar\psi\psi\right>$ 
with the Wilson loop~(\ref{lin}) 
is logarithmically UV divergent in 4d. Rather, the validity of the approximation~(\ref{appr}) 
for a heavy quark has been proven by demonstrating
that, for the Wilson loop given by eq.~(\ref{quadr}), the standard QCD sum-rules result for the 
heavy-quark condensate~\cite{svz} is reproduced correctly. Nevertheless, 
it is instructive to calculate the heavy-quark condensates, $\left<\bar\psi\psi\right>$
and $g\left<\bar\psi\sigma_{\mu\nu}F_{\mu\nu}\psi\right>$, in the case~(\ref{lin}) as well. Firstly, because in the dual Abelian Higgs model [and its 
SU($N$)-generalization] a physical UV cutoff exists, that is the mass of the dual Higgs field, $M_H$. Secondly, this calculation is 
interesting in order to find for the case~(\ref{lin}), in the same way as for the case~(\ref{quadr}), the corresponding critical dimension
of space-time. This can be defined as such a dimension 
$d_c$, that, at $d<d_c$ the corresponding condensate is UV finite, it diverges as $\ln\frac{\Lambda}{m}$ at $d=d_c$, and as $\Lambda^{d-d_c}$
at $d>d_c$. 
Here is a table, which summarizes the values of $d_c$: those for $\left<\bar\psi\psi\right>$ were found in~\cite{4}, whereas those for 
$g\left<\bar\psi\sigma_{\mu\nu}F_{\mu\nu}\psi\right>$ will be found in the present paper.

\vspace{3mm}
\begin{tabular}{||p{38mm}||p{38mm}||p{38mm}||}
\hline
{} & \centerline{$\left<\bar\psi\psi\right>$} & \centerline{$g\left<\bar\psi\sigma_{\mu\nu}F_{\mu\nu}\psi\right>$}\\
\hline
\centerline{case~(\ref{quadr})} & \centerline{6} & \centerline{8}\\
\hline
\centerline{case~(\ref{lin})} & \centerline{4} & \centerline{2}\\ 
\hline
\end{tabular}
\vspace{3mm}

It is further interesting to go beyond approximations~(\ref{quadr}) and~(\ref{lin}) by accounting for the variation of the gauge field,
which is significant at the distances of the order of $T_g$. In this paper, we manage to do so and find, within the stochastic 
vacuum model, an explicit correction to the formula~(\ref{quadr}), in case when ${\cal C}$ is a circle. This allows us to estimate
(at least parametrically) corrections to $\left<\bar\psi\psi\right>$ and $g\left<\bar\psi\sigma_{\mu\nu}F_{\mu\nu}\psi\right>$,
which are produced by the variation of the field. In particular, as far as their signs are concerned, we find that both corrections
are negative, i.e. they diminish the absolute values of the condensates.

The paper is organized as follows. 
In the next section, the mixed condensate will be derived in both cases,~(\ref{quadr}) and~(\ref{lin}). In section~3, these results
will be generalized to finite temperatures. In section~4, corrections emerging due to the variation of the gauge field will be evaluated.
In section~5, a brief summary of the obtained results will be given.

\section{Mixed quark-gluon condensate at zero temperature}

To evaluate the condensate $g\left<\bar\psi\sigma_{\mu\nu}F_{\mu\nu}\psi\right>$, we will use the world-line representation for 
this quantity~\cite{8} and write explicitly the integral over the Grassmann functions $\psi_\mu$'s, which describe spin degrees of freedom
(cf. refs.~\cite{1,2}):
$$
g\left<\bar\psi\sigma_{\mu\nu}F_{\mu\nu}\psi\right>=-\frac{4g^2}{V}\left(m-\gamma_\mu\frac{\partial}{\partial x_\mu(0)}\right)
\int_{\Lambda^{-2}}^{\infty}
dT{\rm e}^{-m^2T}\times$$
$$\times \int\limits_{x_\mu(0)=x_\mu(T)}^{} {\cal D}x_\mu
\int\limits_{\psi_\mu(0)=-\psi_\mu(T)}^{} {\cal D}\psi_\mu
\exp\left[-\int_{0}^{T}d\tau\left(\frac14\dot x_\mu^2+
\frac12\psi_\mu\dot\psi_\mu\right)\right]\times$$
\begin{equation}
\label{1}
\times\int_{0}^{T}d\tau\sigma_{\mu\nu}\sigma_{\lambda\rho}{\,}{\rm tr}\left<F_{\mu\nu}(x(0))F_{\lambda\rho}(x(\tau))\right>
\left<{\cal P}\exp\left[
ig\int_{0}^{T}d\tau\left(A_\mu\dot x_\mu-\psi_\mu\psi_\nu
F_{\mu\nu}\right)\right]\right>.
\end{equation}
Here, $V$ is the four-volume occupied by the system, 
$\Lambda$ stands for the UV momentum cutoff, $\sigma_{\mu\nu}=\frac{1}{4i}[\gamma_\mu,\gamma_\nu]$ with $\gamma_\mu$'s being 
the Euclidean gamma-matrices, the average
$\left<\ldots\right>$ is defined with respect to the gluodynamics action
in the Euclidean space-time, $\frac14 \int d^4 x\left(F^a_{\mu\nu}\right)^2$,
where $a=1,\ldots, N^2-1$ and $F^a_{\mu\nu} =\partial _\mu
A^a_\nu-\partial_\nu A_\mu^a+gf^{abc} A_\mu^bA_\nu^c$ stands for
the Yang-Mills field-strength tensor. Next, 
$A_\mu\equiv A_\mu^a T^a$ with $T^a$'s standing for the 
generators of the SU($N$)-group in the fundamental 
representation, $\left[T^a,T^b\right]=if^{abc}T^c$,
${\rm tr}~ T^aT^b=\frac12\delta^{ab}$. Finally, on the r.h.s. of eq.~(\ref{1}) $4=2(2s+1)$, where $s=1/2$ is the spin of a quark, and 
the factor $(2s+1)$ counts the number of spin degrees of freedom~\cite{1,2}. Rather, an extra factor 2 appears from the expression for
the mixed condensate as a variational derivative with respect to the coupling $g$ (which is made $x$-dependent for a while~\cite{8})
of the averaged quark propagator, which is defined at a closed path: 
$$g\left<\bar\psi\sigma_{\mu\nu}F_{\mu\nu}\psi\right>={\rm tr}{\,}\left<g(x)\sigma_{\mu\nu}F_{\mu\nu}(x)S\left(x,x|A_\mu^a\right)\right>=
2{\,}{\rm tr}{\,}\frac{\delta}{\delta g(0)}\left<S\left(x,x|A_\mu^a\right)\right>,~ {\rm where}$$
$$S\left(x,y|A_\mu^a\right)=(m+\gamma_\mu D_\mu)_{x,y}^{-1},~ D_\mu\equiv\partial_\mu-igA_\mu^aT^a.$$

As in ref.~\cite{4}, by virtue of 
the operator of the area derivative, $\delta/\delta s_{\mu\nu}(x(\tau))$, 
all the gauge-field dependence of this expression can be reduced to that of the Wilson loop $\left<W({\cal C})\right>\equiv
\left<{\rm tr}{\,}{\cal P}\exp\left(ig\int_{0}^{T}
d\tau A_\mu\dot x_\mu\right)\right>$.
Indeed, 
$${\rm tr}\left<F_{\mu\nu}(x(0))F_{\lambda\rho}(x(\tau))\right>
\left<{\cal P}\exp\left[
ig\int_{0}^{T}d\tau\left(A_\mu\dot x_\mu-\psi_\mu\psi_\nu
F_{\mu\nu}\right)\right]\right>=$$
$$=-\frac{4}{g^2}\frac{\delta^2}{\delta s_{\mu\nu}(x(0))\delta s_{\lambda\rho}(x(\tau))}
\exp\left(-2\int_{0}^{T}d\tau\psi_\mu\psi_\nu
\frac{\delta}{\delta s_{\mu\nu}(x(\tau))}\right)\left<W({\cal C})\right>.$$
Further, by splitting 
in the standard way~\cite{2} the coordinate $x_\mu(\tau)$ into the position of a trajectory 
and the relative coordinate, one gets from the empty integration over the position 
the factor $V$, which cancels with $1/V$ on the r.h.s. of eq.~(\ref{1}). The relative coordinate 
will be denoted below as $z_\mu(\tau)$.

\subsection{Stochastic vacuum model}

To start with, let us present a derivation of eq.~(\ref{quadr}) from the stochastic vacuum model. This model
suggests the following parametrization for the nonperturbative part of the 
two-point correlation function of gluonic field strengths:
$$\left<F_{\mu\nu}(x)\Phi_{xx'}F_{\lambda\rho}(x')\Phi_{x'x}\right>=
\frac{\hat 1_{N\times N}}{N}{\cal N}
\Biggl\{(\delta_{\mu\lambda}\delta_{\nu\rho}-\delta_{\mu\rho}\delta_{\nu\lambda})D\left((x-x')^2\right)+$$
$$
+\frac12\left[\partial_\mu^x\left((x-x')_\lambda\delta_{\nu\rho}-(x-x')_\rho\delta_{\nu\lambda}\right)+
\partial_\nu^x\left((x-x')_\rho\delta_{\mu\lambda}-(x-x')_\lambda\delta_{\mu\rho}\right)\right]
D_1\left((x-x')^2\right)\Biggr\}.$$
Here, $\Phi_{xx'}\equiv\frac{1}{N}{\,}{\cal P}{\,}\exp\left(ig\int_{x'}^{x}dz_\mu A_\mu(z)\right)$ is a phase 
factor along the straight line, which goes through $x'$ and $x$, and 
\begin{equation}
\label{norm}
{\cal N}=\frac{\left<F^2\right>}{4n[D(0)+D_1(0)]}
\end{equation}
is the normalization constant. The corresponding expression for the Wilson loop reads (see e.g.~\cite{ad} for the details):
$$\left<W({\cal C})\right>=\exp\Biggl\{-\frac{g^2{\cal N}}{8N}\Biggl[2
\int_{\Sigma({\cal C})}^{}d\sigma_{\mu\nu}(x)
\int_{\Sigma({\cal C})}^{}d\sigma_{\mu\nu}(x')D\left((x-x')^2\right)+$$
\begin{equation}
\label{Wsvm}
+\oint_{{\cal C}}^{}dx_\mu
\oint_{{\cal C}}^{}dx'_\mu G\left((x-x')^2\right)
\Biggr]\Biggr\},~~ G(x^2)\equiv\int_{x^2}^{\infty}dt D_1(t).
\end{equation}
Let us now perform a Taylor expansion of this expression at the  
distances $|x|\lesssim T_g$ of interest. In the leading approximation, we have 
\begin{equation}
\label{0}
G(x^2)\simeq G(0)-x^2D_1(0),~~ D(x^2)\simeq D(0),
\end{equation}
that yields 
$$\oint_{{\cal C}}^{}dx_\mu
\oint_{{\cal C}}^{}dx'_\mu G\left((x-x')^2\right)\simeq 2D_1(0)\Sigma_{\mu\nu}^2,$$    
$$\int_{\Sigma({\cal C})}^{}d\sigma_{\mu\nu}(x)
\int_{\Sigma({\cal C})}^{}d\sigma_{\mu\nu}(x')D\left((x-x')^2\right)\simeq D(0)\Sigma_{\mu\nu}^2.$$
Taking into account the value of the normalization constant ${\cal N}$, eq.~(\ref{norm}), we arrive in this approximation at 
eq.~(\ref{quadr}). In section~4, by finding the next term of the Taylor expansion, we will obtain corrections to eq.~(\ref{quadr}) and 
to the quark condensates produced by the variation of the gauge field.

Further, 
the square of the tensor area in the exponent in eq.~(\ref{quadr}) can be linearized by using the 
Hubbard-Stratonovich transformation with a constant-valued antisymmetric-tensor field $B_{\mu\nu}$~\cite{4}
$$
N\exp\left(-C\Sigma_{\mu\nu}^2\right)=\frac{N}{(8\pi C)^{n/2}}
\Biggl(\prod\limits_{\mu<\nu}^{}\int_{-\infty}^{+\infty}dB_{\mu\nu}
{\rm e}^{-\frac{B_{\mu\nu}^2}{8C}}\Biggr)
\exp\left(-\frac{i}{2}
B_{\mu\nu}\Sigma_{\mu\nu}\right),$$
where $n=\frac{d^2-d}{2}$ is a half of the number of off-diagonal components of $B_{\mu\nu}$ in $d$ dimensions.
The operator $\delta/\delta s_{\mu\nu}$ can now readily be applied, since, due to the Stokes' theorem, 
$B_{\mu\nu}\Sigma_{\mu\nu}=\int_{\Sigma({\cal C})}^{}d\sigma_{\mu\nu}B_{\mu\nu}$. In particular, since the field $B_{\mu\nu}$ is space-time
independent, the operator $\partial/\partial x_\mu(0)$ in eq.~(\ref{1}) yields zero.
The world-line integral becomes the one in a constant Abelian field, and the result is nothing but 
the Euler-Heisenberg-Schwinger Lagrangian (see e.g.~\cite{2}):
$$
\int\limits_{z_\mu(0)=z_\mu(T)}^{} {\cal D}z_\mu
\int\limits_{\psi_\mu(0)=-\psi_\mu(T)}^{} {\cal D}\psi_\mu\times$$
$$\times\exp\left[-\int_{0}^{T}d\tau\left(\frac14\dot z_\mu^2+
\frac12\psi_\mu\dot\psi_\mu+\frac{i}{2}B_{\mu\nu}z_\mu\dot z_\nu
-iB_{\mu\nu}\psi_\mu\psi_\nu\right)\right]-
\frac{1}{(4\pi T)^{d/2}}=$$
$$=\frac{1}{(4\pi T)^{d/2}}\left[T^2ab\cot(aT)\coth(bT)-1\right].
$$
Here, we have in the standard way subtracted the part of the integral without the field, in order to respect the obvious normalization condition
$\left.g\left<\bar\psi\sigma_{\mu\nu}F_{\mu\nu}\psi\right>\right|_{A_\mu^a=0}=0$.
We have also used the standard notations~\cite{2}
$$a^2=\frac12\left[{\bf E}^2-{\bf H}^2+
\sqrt{\left({\bf E}^2-{\bf H}^2\right)^2+4({\bf E}
\cdot{\bf H})^2}{\,}\right],$$ 
$$b^2=\frac12\left[-\left({\bf E}^2-{\bf H}^2\right)+
\sqrt{\left({\bf E}^2-{\bf H}^2\right)^2+
4({\bf E}\cdot{\bf H})^2}{\,}\right]$$
with ${\bf E}=i\left(B_{41},B_{42},B_{43}\right)$ and ${\bf H}=\left(B_{23},-B_{13},B_{12}\right)$ 
being the electric and magnetic fields, which correspond to the field-strength 
tensor $B_{\mu\nu}$ ($B_{ij}=\varepsilon_{ijk}H_k$, $B_{4i}=-iE_i$).

Next, due to the factor ${\rm e}^{-m^2T}$, 
this expression can be expanded in powers of $T$:
$$T^2ab\cot(aT)\coth(bT)-1=$$
\begin{equation}
\label{approx}
=\frac{T^2}{3}\left(b^2-a^2\right)+
{\cal O}\left(T^4({\bf E}\cdot{\bf H})^2\right)=\frac{T^2}{3}\sum\limits_{\alpha<\beta}^{}B_{\alpha\beta}^2+{\cal O}
\left(T^4\left(B_{\mu\nu}^2\right)^2\right).
\end{equation}
Note that, in terms of the fermionic determinant, this approximation corresponds to retaining only the diagram
with two external lines of the field $B_{\mu\nu}$ in the expansion of the logarithm. The parameter of the expansion can 
actually be estimated accurately. To do the estimate, let us set $d\sim 1$, having in mind the values of $d_c$ given by the 
table in section~1. Then, since $T\sim m^{-2}$ and $B_{\mu\nu}^2\sim 
g^2\left<F^2\right>/N$, the parameter of the expansion is ${\cal O}\left(
\frac{g^2\left<F^2\right>}{Nm^4}\right)$, i.e. the heavy-quark limit is implied in the sense of the inequality
\begin{equation}
\label{heavy}
\frac{g^2\left<F^2\right>}{Nm^4}\lesssim 1.
\end{equation}
For $N=3$, the expansion~(\ref{approx}) breaks down for $u$, $d$ and $s$ quarks, but holds for $c$, $b$ and $t$ quarks under study. 
Note also that the parameter of the expansion is ${\cal O}\left(N^0\right)$ at large $N$.

As follows from eq.~(\ref{approx}), the adopted approximation reduces
the range of the $B_{\mu\nu}$-integration to 
the interval $(-1/T, 1/T)\sim(-m^2, m^2)$. However, the characteristic values of $\sqrt{B_{\mu\nu}^2}$, being of the order of 
${\cal O}\left(\sqrt{g^2\left<F^2\right>/N}\right)$, are 
smaller than
$m^2$, as long as the expansion~(\ref{approx}) converges. 
This fact enables one to retain the infinite range of the $B_{\mu\nu}$-integration with the 
exponential accuracy.  Apparently, this is true only for the case~(\ref{quadr})
under study, when the measure of the $B_{\mu\nu}$-integration falls off so rapidly. Rather, in case~(\ref{lin}), 
which will be considered in the next subsection, the measure will be shown to fall off only as some power 
of $B_{\mu\nu}^2$, and the infinite range of integration over $B_{\mu\nu}$ cannot be retained anymore.

The appearing integral over $B_{\mu\nu}$ can then be readily calculated:
\begin{equation}
\label{Bint}
\frac{1}{(8\pi C)^{n/2}}
\Biggl(\prod\limits_{\mu<\nu}^{}\int_{-\infty}^{+\infty}dB_{\mu\nu}
{\rm e}^{-\frac{B_{\mu\nu}^2}{8C}}\Biggr)B_{\mu\nu}B_{\lambda\rho}
\sum\limits_{\alpha<\beta}^{}B_{\alpha\beta}^2=32\left(\frac{n}{2}+1\right)C^2
(\delta_{\mu\lambda}\delta_{\nu\rho}-\delta_{\mu\rho}\delta_{\nu\lambda}).
\end{equation}
Bringing all the factors together, taking the limit $\Lambda\to\infty$, and noticing that $\sigma_{\mu\nu}^2=\frac{n}{2}$, we have 
\begin{equation}
\label{condensate}
g\left<\bar\psi\sigma_{\mu\nu}F_{\mu\nu}\psi\right>=-\frac{128n\left(\frac{n}{2}+1\right)N}{3(4\pi)^{d/2}}mC^2\int_{0}^{\infty}
\frac{dT{\rm e}^{-m^2T}}{T^{\frac{d}{2}-3}}.
\end{equation}
For the case $d<8$, where the integral is convergent, we obtain the final result (valid at $d\ge2$):
\begin{equation}
\label{finalsvm}
g\left<\bar\psi\sigma_{\mu\nu}F_{\mu\nu}\psi\right>=
-\frac{\left(\frac{n}{2}+1\right)\Gamma\left(4-\frac{d}{2}\right)}{6(4\pi)^{d/2}nN}
\left(g^2\left<F^2\right>\right)^2m^{d-7},
\end{equation}
where ``$\Gamma$'' henceforth stands for the Gamma-function.
As well as the standard heavy-quark condensate $\left<\bar\psi\psi\right>$, this expression scales at large $N$ as ${\cal O}(N)$.
At $d<6$, when $\left<\bar\psi\psi\right>$ is also UV finite~\cite{4}, we obtain for the ratio 
of the two condensates~\footnote{As in ref.~\cite{8}, this ratio is defined with the factor 2, in order to bring it in accordance with 
an analogous quantity used in QCD sum rules. The reason is due to the different definitions of $\sigma_{\mu\nu}$: 
$\sigma_{\mu\nu}=\frac{1}{2i}[\gamma_\mu,\gamma_\nu]$ in QCD sum rules, while $\sigma_{\mu\nu}=\frac{1}{4i}[\gamma_\mu,\gamma_\nu]$
in~\cite{8} and in the present paper.}
\begin{equation}
\label{our}
m_0^2\equiv2\frac{g\left<\bar\psi\sigma_{\mu\nu}F_{\mu\nu}\psi\right>}{\left<\bar\psi\psi\right>}=
\frac{\left(\frac{n}{2}+1\right)\left(3-\frac{d}{2}\right)}{nN}\frac{g^2\left<F^2\right>}{m^2}.
\end{equation}
In particular, 
$$
\left.m_0^2\right|_{d=4}=\frac23\frac{g^2\left<F^2\right>}{Nm^2}.
$$

\subsection{Dual superconductor model}

In this case, one needs an additional integration over an ``einbein'' parameter, in order 
to get rid of the square root in the exponent in eq.~(\ref{lin}):
$$
\left<W({\cal C})\right>=N\int_{0}^{\infty}\frac{d\lambda}{\sqrt{\pi\lambda}}
\exp\left(-\lambda-\frac{\sigma^2\Sigma_{\mu\nu}^2}{8\lambda}\right).$$
Using further again the Hubbard-Stratonovich transformation and integrating over $\lambda$, one gets 
the following expression for the Wilson loop~\cite{4}:
$$\left<W({\cal C})\right>=
\frac{\Gamma\left(\frac{n+1}{2}\right)N}{\pi^{\frac{n+1}{2}}\sigma^n}\left(\prod\limits_{\mu<\nu}^{}
\int_{-\infty}^{+\infty}dB_{\mu\nu}\right)
\frac{\exp\left(-\frac{i}{2}
B_{\mu\nu}\Sigma_{\mu\nu}\right)}{\left(1+\frac{1}{2\sigma^2}B_{\mu\nu}^2\right)^{\frac{n+1}{2}}}.
$$
The weight of the $B_{\mu\nu}$-integration in this case is therefore only polynomial and not Gaussian. Let us adopt again the 
approximation~(\ref{approx}), where the expansion parameter now becomes $\sigma/m^2$, that establishes an upper bound on the value of $\sigma$.
Taking for instance the minimal value of $m$, $m_c=1.1{\,}{\rm GeV}$, and parametrizing $\sigma=\lambda\sigma_{\infty}$, we obtain $\lambda<6.7$.

The $B_{\mu\nu}$-integration then takes the form [cf. eq.~(\ref{Bint})]:
$$
\left(\prod\limits_{\mu<\nu}^{}
\int_{-1/T}^{1/T}dB_{\mu\nu}\right)\frac{B_{\mu\nu}B_{\lambda\rho}}{\left(1+\frac{1}{2\sigma^2}B_{\mu\nu}^2\right)^{\frac{n+1}{2}}}
\sum\limits_{\alpha<\beta}^{}B_{\alpha\beta}^2=$$
$$=\frac{\pi^{n/2}\sigma^{n+4}}{\Gamma\left(\frac{n}{2}+1\right)}\int_{0}^{(\sigma T)^{-1}}
dx\frac{x^{n+3}}{(1+x^2)^{\frac{n+1}{2}}}\cdot(\delta_{\mu\lambda}\delta_{\nu\rho}-\delta_{\mu\rho}\delta_{\nu\lambda}).$$
Denoting $t=m^2T$, we arrive at the following intermediate result:
$$
g\left<\bar\psi\sigma_{\mu\nu}F_{\mu\nu}\psi\right>=-
\frac{2^{3-d}N\Gamma\left(\frac{n+1}{2}\right)\sigma^4m^{d-7}}{3\Gamma\left(\frac{n}{2}\right)
\pi^{\frac{d+1}{2}}}\int\limits_{(m/\Lambda)^2}^{\infty}dtt^{3-\frac{d}{2}}{\rm e}^{-t}
\int\limits_{0}^{\frac{m^2}{\sigma t}}\frac{dxx^{n+3}}{\left(1+x^2\right)^{\frac{n+1}{2}}}.$$
Clearly, the integral over $x$ here receives the dominant contribution 
from the region of $t$ around $(m/\Lambda)^2$. For such $t$'s,
the $x$-integral approximately equals $\frac13\left(\frac{m^2}{\sigma t}\right)^3$, and we arrive at the following result: 
at $d=2$, the condensate diverges logarithmically in the UV region, namely 
\begin{equation}
\label{finalahm}
\left.g\left<\bar\psi\sigma_{\mu\nu}F_{\mu\nu}\psi\right>\right|_{d=2}=-\frac{4N\sigma m}{(3\pi)^2}\ln\frac{\Lambda}{m},
\end{equation}
while, at $d>2$ it diverges as $\Lambda^{d-2}$, i.e. polynomially:
\begin{equation}
\label{finalahm1}
\left.g\left<\bar\psi\sigma_{\mu\nu}F_{\mu\nu}\psi\right>\right|_{d>2}=-\frac{2^{3-d}\Gamma\left(\frac{n+1}{2}\right)N\sigma 
m\Lambda^{d-2}}{9\left(\frac{d}{2}-1\right)\pi^{\frac{d+1}{2}}\Gamma\left(\frac{n}{2}\right)}.
\end{equation}

Setting for $\Lambda$ the mass of the dual Higgs boson, $M_H$, we can obtain a lower bound for the value of the Landau-Ginzburg parameter 
$\kappa\equiv M_H/M_V$, where $M_V$ is the mass of the dual vector boson. Indeed, the condition $\Lambda\gg m$ yields 
\begin{equation}
\label{kappa}
\ln\kappa\gg\ln\frac{m}{M_V}.
\end{equation}  
On the other hand, it can be shown~\cite{corr} that $M_V^{-1}$ is the distance,
at which the two-point correlation function of $F_{\mu\nu}$'s in the dual Abelian Higgs model, or its SU($N$)-generalization, 
falls off. Therefore, $M_V$ should be of the order of $T_g^{-1}$. For $c$ and $b$ quarks, this value of $M_V$ leads to the 
condition $\ln\kappa\gg 1$, that is the standard condition of the London limit in the type-II (dual) superconductor~\cite{lp}.
Rather, for $t$ quark, whose mass is about 170~GeV, this condition should be replaced by $\ln\kappa\gg 5.1$. Therefore, it makes sense to 
speak about the heavy-quark condensates in this model, provided the condition of the London limit is fulfilled in a certain, $m$-dependent, way.

\section{Finite-temperature generalizations}

In this section, we will generalize the above-obtained 4d-results to the case of finite temperatures. In case~(\ref{quadr}), 
we will adopt the QCD-value for the temperature of dimensional reduction, $T_{\rm d.r.}\simeq 2T_c$ (see e.g.~\cite{110}).
At temperatures smaller than $T_{\rm d.r.}$, it is important to account for the 
antiperiodic boundary conditions for quarks, that can be done upon the multiplication of the zero-temperature heat kernel by the 
factor~\cite{81} $\left[1+2\sum\limits_{n=1}^{\infty}(-1)^n{\rm e}^{-\frac{\beta^2n^2}{4T}}\right]$. Here, $\beta$ is the inverse temperature, and 
$n$ is the number of a Matsubara mode. Inserting this factor into eq.~(\ref{condensate}) and calculating the integral over the proper time,
we have for the case~(\ref{quadr}):
$$
g\left<\bar\psi\sigma_{\mu\nu}F_{\mu\nu}\psi\right>(T)\equiv G(T)=$$
$$=G(0)\left[1+(\beta m)^2\sum\limits_{n=1}^{\infty}(-1)^nn^2K_2(\beta mn)\right]
\simeq G(0)\left[1-\sqrt{\pi/2}(\beta m)^{3/2}{\rm e}^{-\beta m}\right],$$
where $G(0)=-\frac{1}{144\pi^2}\frac{(g^2\left<F^2\right>)^2}{Nm^3}$.
This result is valid at $T<T_c$, where, for heavy quarks, $\beta m\gg 1$. At $T=T_c$, the chromoelectric condensate evaporates, and one should 
substitute $2g^2\left<(B_\mu^a)^2\right>$ instead of $g^2\left<F^2\right>$. Note that the lattice data~\cite{Temp} also confirm that, at $T=T_c$,
$g^2\left<F^2\right>$ drops by a factor of the order of 2. With this substitution, the obtained result is valid up to $T_{\rm d.r.}$, since 
$\beta m>1$ up to this temperature. At $T>T_{\rm d.r.}$, the sum over Matsubara frequences disappears, and one is left with a 3d theory with
the choromomagnetic condensate. The mixed condensate in this theory is given by eq.~(\ref{finalsvm}) at $d=3$. 
As follows from the elementary dimensional analysis, which relates fields and couplings in this theory to those in the original 4d one,
the mixed condensate in the original QCD at $T>T_{\rm d.r.}$ stems from this result upon the multiplication by $T$. It reads 
$$G(T)=-\frac{5T}{96\pi N}\frac{\left(g^2\left<(B_\mu^a)^2\right>\right)^2}{m^4}.$$

Let us now consider the dual Abelian-Higgs--type theory at finite temperature. Since $M_H(T)\propto\sqrt{\sigma(T)}$, where 
$\sigma(T)\sim\left(1-\frac{T}{T_c}\right)^\nu$ at $T\to T_c-0$, we have 
$$
\frac{M_H(T)}{m}=\frac{M_H(0)}{m}\left(1-\frac{T}{T_c}\right)^{\nu/2}\gg1.$$
Since $M_H(0)\equiv\kappa M_V(0)$, we obtain a new condition for $\kappa$ [cf. eq.~(\ref{kappa})]:
\begin{equation}
\label{kappaT}
\ln\kappa\gg\ln\frac{m}{M_V(0)}-\frac{\nu}{2}\ln\left(1-\frac{T}{T_c}\right).
\end{equation}
Here, $\nu$ should be set 1/3 and 0.63~\cite{ep} for the 1-st and the 2-nd order phase transition, respectively.
The obtained formula defines the law, according to which the constraint imposed on $\kappa$ should become more
severe when $T$ approaches $T_c$. Provided this constraint is obeyed, we then have at $T<T_c$:
$$
G(T)=G(0)\left[1+\frac{8mT}{M_H^2(T)}\sum\limits_{n=1}^{\infty}\frac{(-1)^n}{n}K_1(\beta mn)\right]
\simeq G(0)\left[1-4\sqrt{2\pi}\frac{(mT^3)^{1/2}}{M_H^2(T)}{\rm e}^{-\beta m}\right].$$
Here, $G(0)=-\frac{5NmM_H^2(0)\sigma}{96\pi^2}$, and we have assumed that, as well as in QCD, $T_c\ll m$.

\section{Accounting for the variation of the gauge field within 
the stochastic vacuum model}
In this section, we will estimate corrections to $\left<\bar\psi\psi\right>$ and $m_0^2$ at $T=0$, which can be obtained
within the stochastic vacuum model by treating the gauge fields as varying, in the leading approximation.
To this end, we need to find the next term in the Taylor expansion of
eq.~(\ref{Wsvm}). This is only possible if we specify the form of the functions $D$ and $D_1$ at the 
distances $|x|\lesssim T_g$ of interest. In ref.~\cite{chir}, it has been argued that, at such distances, 
these functions should have a Gaussian shape. This follows for instance from the formula $\left.\frac{dD(x^2)}{dx^2}\right|_{x=0}\propto gf^{abc}
\left<F_{\mu\nu}^aF_{\nu\lambda}^bF_{\lambda\mu}^c\right>$, which, due to the finiteness of the triple condensate, 
means that $D$ should be an analytic function of $x^2$ at short distances. We therefore take the functions 
$D$ and $D_1$ in the form $D(x^2)=D(0){\rm e}^{-x^2/T_g^2}$, $D_1(x^2)=D_1(0){\rm e}^{-x^2/T_g^2}$.
The extra terms, one then gets in eq.~(\ref{0}), read 
$\Delta G(x^2)=\frac{D_1(0)}{2T_g^2}(x^2)^2$, $\Delta D(x^2)=-\frac{D(0)}{T_g^2}x^2$. 
To estimate the effect, which these corrections produce on the condensates,
we will assume for simplicity that ${\cal C}$ is a circle of the radius $R$.
We then obtain very similar expressions for the 
corresponding integrals: 
$$\oint_{{\cal C}}^{}dx_\mu
\oint_{{\cal C}}^{}dx'_\mu\Delta G\left((x-x')^2\right)=-16\pi^2\frac{D_1(0)}{T_g^2}R^6,$$
$$2\int_{\Sigma({\cal C})}^{}d\sigma_{\mu\nu}(x)
\int_{\Sigma({\cal C})}^{}d\sigma_{\mu\nu}(x')\Delta D\left((x-x')^2\right)=-16\pi^2\frac{D(0)}{T_g^2}R^6,$$ 
where in the last case 
we have used the formula $d\sigma_{\mu\nu}(x)=\frac12(x_\mu dx_\nu-x_\nu dx_\mu)$. Combining together the leading and the 
next-order results, we obtain 
$$
\left<W({\cal C})\right>\simeq N\exp\left[-C\Sigma_{\mu\nu}^2+\alpha(\Sigma_{\mu\nu}^2)^{3/2}\right],~ 
\alpha=\frac{g^2\left<F^2\right>}{2^{3/2}\pi N(d^2-d)T_g^2},$$
where we have used that $R^6=\frac{(\Sigma_{\mu\nu}^2)^{3/2}}{2^{3/2}\pi^3}$. The ratio of the correction to the absolute value of the leading term 
is $\frac{\alpha(\Sigma_{\mu\nu}^2)^{3/2}}{C\Sigma_{\mu\nu}^2}\sim\frac{\sqrt{\Sigma_{\mu\nu}^2}}{T_g^2}$.
This means that the correction is anyway small
as long as the size of ${\cal C}$, that is the quark's wavelength $1/m$, is smaller than $T_g$ -- the approximation we work in.
To perform the calculations, we will, however, need the validity of a more fine inequality, which will be derived in a moment.

Apparently, the correction can be analysed if we get rid of the power 3/2. This can be done by noticing that 
${\rm e}^{\alpha (\Sigma_{\mu\nu}^2)^{3/2}}$ is nothing but the saddle-point value of the integral 
$c\int_{0}^{\infty}d\lambda\exp\left(-\frac{4}{27\alpha^2}\lambda^3+\lambda\Sigma_{\mu\nu}^2\right)$. 
The normalization constant $c$ here is fixed by the condition $1=c\int_{0}^{\infty}d\lambda{\rm e}^{-\frac{4}{27\alpha^2}\lambda^3}$, which
yields $c=\frac{(2/\alpha)^{2/3}}{\Gamma(1/3)}$. The final expression for the Wilson loop therefore reads
$$\left<W({\cal C})\right>\simeq\frac{N(2/\alpha)^{2/3}}{\Gamma(1/3)}\int_{0}^{\infty}d\lambda\exp\left\{-\left[
\frac{4}{27\alpha^2}\lambda^3+(C-\lambda)\Sigma_{\mu\nu}^2\right]\right\}.$$
According to this formula, the requirement for the next-order result to be small with respect to the leading one implies that the characteristic 
values of $\lambda$ are to be much smaller than $C$. Parametrically, these characteristic values are of the order of $\alpha^{2/3}$, that
leads to the condition
$\frac{g^2\left<F^2\right>T_g^4}{N(d^2-d)}\gg 1$. This 
inequality apparently breaks down at large $d$, that does not matter anyway, since the condensates diverge at $d\ge d_c$. At $d<d_c$, we therefore
have, together with the heavy-quark limit condition~(\ref{heavy}), the following sequence of inequalities:
$$\frac{1}{m}\lesssim\left(\frac{N}{g^2\left<F^2\right>}\right)^{1/4}\ll T_g.$$
It clearly tells us that the quark wavelength is smaller than that of the gauge field, 
which itself is much smaller than the vacuum correlation length. The first inequality means that, inside ${\cal C}$, the 
gauge field is effectively constant, while the second one means that the field is varying significantly at the 
scale of the vacuum correlation length.

The last step in evaluating the desired corrections to the condensates essentially amounts to substitute 
$C\to C-\lambda$, $C^2\to C^2-2C\lambda$ and integrate over $\lambda$. We obtain 
\begin{equation}
\label{cor1}
\left<\bar\psi\psi\right>_{\rm new}=\left<\bar\psi\psi\right>(1-\xi),~ 
g\left<\bar\psi\sigma_{\mu\nu}F_{\mu\nu}\psi\right>_{\rm new}=g\left<\bar\psi\sigma_{\mu\nu}F_{\mu\nu}\psi\right>(1-2\xi),~
m_{0{\,}{\rm new}}^2=m_0^2(1-\xi),
\end{equation}
where 
\begin{equation}
\label{cor2}
\xi=\frac{(2/\alpha)^{2/3}}{\Gamma(1/3)C}\int_{0}^{\infty}d\lambda\lambda
\exp\left(-\frac{4}{27\alpha^2}\lambda^3\right)=\frac{3\cdot 2^{4/3}\cdot\Gamma(2/3)}{\pi^{2/3}\cdot\Gamma(1/3)}
\left[\frac{N(d^2-d)}{g^2\left<F^2\right>T_g^4}\right]^{1/3},
\end{equation}
and $\left<\bar\psi\psi\right>=-\frac{\Gamma\left(3-\frac{d}{2}\right)m^{d-5}}{3(4\pi)^{d/2}}g^2\left<F^2\right>$ is the 
value of the heavy-quark condensate at $2\le d<6$~\cite{4} without the correction. This result gives an idea of how the 
variation of the gauge field affects the condensates in the leading approximation. In particular, we see that the corrections 
diminish the absolute values of both condensates, as well as the value of $m_0^2$.

\section{Conclusions}
We have calculated the mixed quark-gluon condensate, in the heavy-quark limit, at zero and finite temperatures. This has been done in the models 
of stochastic vacuum and dual superconductor, where the nonperturbative parts of the Wilson loop at the distances smaller than the 
vacuum correlation length are given by eqs.~(\ref{quadr}) and (\ref{lin}), respectively. In the stochastic vacuum model, the condensate is UV 
divergent at $d\ge 8$, while in the dual superconductor it diverges right at $d\ge 2$. This divergence can be regularized by attributing 
to the UV cutoff a physical meaning of the inverse thickness of a ``short string'', that is nothing but the mass of the dual Higgs field. 
We have also found the constraints, which the Landau-Ginzburg parameter
should obey at zero and finite temperatures, in order for such an interpretation to be possible [cf. eqs.~(\ref{kappa}) and~(\ref{kappaT})]. 
At zero temperature, the obtained expressions for the 
condensate are given by eqs.~(\ref{finalsvm}), (\ref{finalahm}), and (\ref{finalahm1}), while the finite-temperature 
generalizations are discussed in section~3.

We have further evaluated corrections to 
$\left<\bar\psi\psi\right>$
and $g\left<\bar\psi\sigma_{\mu\nu}F_{\mu\nu}\psi\right>$, which appear within the stochastic vacuum model, due to the 
variation of the gauge field at the scale of the vacuum correlation length. 
To this end, we have taken into account that, at the distances smaller than $T_g$, 
the two-point correlation function of gluonic field strengths has a Gaussian shape.
The calculation has been done in the most simple case, namely when the quark trajectory is a circle.
Despite this simplification, the obtained formulae, eqs.~(\ref{cor1}) and~(\ref{cor2}), yield 
the parametric form of the corrections and show 
that these diminish the absolute values of the condensates, as well as the ratio of the mixed condensate to the standard one.

The world-line approach to the calculation of heavy-quark condensates, 
proposed in ref.~\cite{4} and in the present paper, has several positive features. First of all, as we have seen, it allows to calculate
the condensates in various models of the QCD vacuum, where a certain nonperturbative $q\bar q$-potential at small distances is generated.
In particular, the stochastic vacuum model, where this potential is quadratic, reproduces the 4d-value
of $\left<\bar\psi\psi\right>$, which stems from QCD sum rules. We have also seen that the proposed approach naturally allows
for the finite-temperature generalizations. Since the influence of thermal effects on the condensates is an important problem,
it would be very interesting to have the lattice data on $m_0^2(T)$ not only in the chiral
limit~\cite{L}, but also in the heavy-quark one, and to compare them with our results. We have further seen that the world-line approach
is appropriate for a systematic analysis of corrections to the condensates, which originate from the variability of the 
gauge field inside the quark trajectory. Last but not least, our approach can also be used for the calculation of the higher 
mixed heavy-quark--gluon condensates of the form $g^n\left<\bar\psi(\sigma_{\mu\nu}F_{\mu\nu})^n\psi\right>$, where $n\ge 2$ is an 
integer.

Note in conclusion that recently~\cite{chir1}, $\left<\bar\psi\psi\right>$ has been obtained as a function of $m/T$, 
by doing calculations on the gravity side of the AdS/CFT correspondence. The large-$m$ part of that result qualitatively 
agrees with the $1/m$ fall-off, which is predicted by the stochastic vacuum model and QCD sum rules and disagrees 
with the behavior $m\ln\frac{\Lambda}{m}$, which stems from the dual superconductor model~\cite{4}. It would be 
very interesting to have analogous results for $g\left<\bar\psi\sigma_{\mu\nu}F_{\mu\nu}\psi\right>$ as well, and to compare those
with the predictions of the present paper.

\acknowledgments
I am grateful to D.~Ebert, J.~Erdmenger, H.~Gies, H.-J.~Pirner, M.~G.~Schmidt, C.~Schubert, and Yu.~A.~Simonov for useful discussions.
I would also like to thank the Alexander~von~Humboldt Foundation for the financial support and the
staff at the Institute of Physics of the Humboldt University of Berlin for the kind hospitality.


\begin{thebibliography}{99}



\bibitem{5}
B.~L.~Ioffe,
Nucl.\ Phys.\ B {\bf 188}, 317 (1981)
[Erratum-ibid.\ B {\bf 191}, 591 (1981)];
V.~M.~Belyaev and B.~L.~Ioffe,
Sov.\ Phys.\ JETP {\bf 56}, 493 (1982)
[Zh.\ Eksp.\ Teor.\ Fiz.\  {\bf 83}, 876 (1982)];
L.~J.~Reinders, H.~R.~Rubinstein and S.~Yazaki,
Phys.\ Lett.\ B {\bf 120}, 209 (1983)
[Erratum-ibid.\ B {\bf 122}, 487 (1983)];
A.~A.~Ovchinnikov and A.~A.~Pivovarov,
Sov.\ J.\ Nucl.\ Phys.\  {\bf 48}, 721 (1988)
[Yad.\ Fiz.\  {\bf 48}, 1135 (1988)];
S.~Narison,
Phys.\ Lett.\ B {\bf 210}, 238 (1988);
H.~G.~Dosch, M.~Jamin and S.~Narison,
Phys.\ Lett.\ B {\bf 220}, 251 (1989);
H.~G.~Dosch and S.~Narison,
Phys.\ Lett.\ B {\bf 417}, 173 (1998)
[arXiv:hep-ph/9709215].

\bibitem{6}
S.~Narison,
``Light and heavy quark masses, flavor breaking of chiral condensates,  meson
weak leptonic decay constants in QCD,'' 
(Adapted from a chapter of '{\it QCD as a Theory of Hadrons (from partons to confinement)}', 
Cambridge Monogr. Part. Phys. Nucl. Phys. Cosmol. 17: 1, 2004)
[arXiv:hep-ph/0202200].

\bibitem{7}
M.~Kremer and G.~Schierholz,
Phys.\ Lett.\ B {\bf 194}, 283 (1987);
T.~Doi, N.~Ishii, M.~Oka and H.~Suganuma,
Phys.\ Rev.\ D {\bf 67}, 054504 (2003)
[arXiv:hep-lat/0211039];
T.~W.~Chiu and T.~H.~Hsieh,
Nucl.\ Phys.\ B {\bf 673}, 217 (2003)
[arXiv:hep-lat/0305016].


\bibitem{8}
A.~Di Giacomo and Yu.~A.~Simonov,
Phys.\ Lett.\ B {\bf 595}, 368 (2004)
[arXiv:hep-ph/0404044].

\bibitem{90}
H.~G.~Dosch, 
Phys.\ Lett.\ B {\bf 190}, 177 (1987);
Yu.~A.~Simonov, 
Nucl.\ Phys.\ B {\bf 307}, 512 (1988).

\bibitem{ds}
H.~G.~Dosch and Yu.~A.~Simonov, Phys.\ Lett.\ B {\bf 205}, 339 (1988).


\bibitem{110}
A.~Di~Giacomo, H.~G.~Dosch, V.~I.~Shevchenko and Yu.~A.~Simonov, 
Phys.\ Rept.\  {\bf 372}, 319 (2002) 
[arXiv:hep-ph/0007223].


\bibitem{4}
D.~Antonov,
JHEP {\bf 10}, 030 (2003)
[arXiv:hep-ph/0308026].

\bibitem{1}
Z.~Bern and D.~A.~Kosower,
Phys.\ Rev.\ Lett.\  {\bf 66}, 1669 (1991); 
Nucl.\ Phys.\ B {\bf 379}, 451 (1992);
M.~J.~Strassler, 
Nucl.\ Phys.\ B {\bf 385}, 145 (1992)
[arXiv:hep-ph/9205205];
M.~G.~Schmidt and C.~Schubert, 
Phys.\ Lett.\ B {\bf 318}, 438 (1993) 
[arXiv:hep-th/9309055]; 
ibid. B {\bf 331}, 69 (1994)
[arXiv:hep-th/9403158].


\bibitem{2}
C.~Schubert, 
Phys.\ Rept.\  {\bf 355}, 73 (2001)  
[arXiv:hep-th/0101036].


\bibitem{3}
H.~Gies, J.~Sanchez-Guillen and R.~A.~Vazquez,
``Quantum effective actions from nonperturbative worldline dynamics,''
arXiv:hep-th/0505275.



\bibitem{11}
M.~Campostrini, A.~Di Giacomo and G.~Mussardo,
Z.\ Phys.\ C {\bf 25}, 173 (1984);
A.~Di~Giacomo and H.~Panagopoulos, Phys.\ Lett.\ B {\bf 285}, 133 (1992).


\bibitem{12}
A.~Di~Giacomo, {\it Nonperturbative QCD}, in ``Lisbon 1999, QCD: Perturbative or nonperturbative'', 
p.p. 1-29  [arXiv:hep-lat/9912016]; 
{\it Topics in nonperturbative QCD}, in 
{\it Czech. J. Phys.} {\bf 51}, B9 (2001) 
[arXiv:hep-lat/0012013]; {\it QCD vacuum and confinement}, 
in ``Campos do Jordao 2002, New states of matter in hadronic interactions'', p.p. 168-190 
[arXiv:hep-lat/0204001].


\bibitem{polyakov}
A.~M.~Polyakov,
Nucl.\ Phys.\ B {\bf 120}, 429 (1977).


\bibitem{ms}
S.~Maedan and T.~Suzuki,
Prog.\ Theor.\ Phys.\  {\bf 81}, 229 (1989).

\bibitem{a}
D.~Antonov,
Phys.\ Lett.\ B {\bf 543}, 53 (2002)
[arXiv:hep-th/0207092].

\bibitem{pNR}
For a review see: N.~Brambilla, A.~Pineda, J.~Soto and A.~Vairo,
``Effective field theories for heavy quarkonium,''
arXiv:hep-ph/0410047, part VI, C.

\bibitem{tach}
S.~J.~Huber, M.~Reuter and M.~G.~Schmidt,
Phys.\ Lett.\ B {\bf 462}, 158 (1999)
[arXiv:hep-ph/9906358].

\bibitem{sommer}
S.~Necco and R.~Sommer,
Nucl.\ Phys.\ B {\bf 622}, 328 (2002)
[arXiv:hep-lat/0108008]; 
Phys.\ Lett.\ B {\bf 523}, 135 (2001)
[arXiv:hep-ph/0109093].


\bibitem{svz}
M.~A.~Shifman, A.~I.~Vainshtein and V.~I.~Zakharov, 
Nucl.\ Phys.\ B {\bf 147}, 385 (1979).



\bibitem{ad}
D.~Antonov and A.~Di Giacomo,
JHEP {\bf 03}, 017 (2005)
[arXiv:hep-th/0501065].



\bibitem{16}
See e.g. D.~Antonov, D.~Ebert and Yu.~A.~Simonov,
Mod.\ Phys.\ Lett.\ A {\bf 11}, 1905 (1996)
[arXiv:hep-th/9605086].




\bibitem{corr}
D.~Antonov,
Mod.\ Phys.\ Lett.\ A {\bf 13}, 659 (1998)
[arXiv:hep-th/9710144];
M.~Baker, N.~Brambilla, H.~G.~Dosch and A.~Vairo,
Phys.\ Rev.\ D {\bf 58}, 034010 (1998)
[arXiv:hep-ph/9802273];
D.~Antonov and D.~Ebert,
Eur.\ Phys.\ J.\ C {\bf 8}, 343 (1999)
[arXiv:hep-th/9806153];
U.~Ellwanger,
Eur.\ Phys.\ J.\ C {\bf 7}, 673 (1999)
[arXiv:hep-ph/9807380];
D.~Antonov,
JHEP {\bf 07}, 055 (2000)
[arXiv:hep-ph/0006156].

\bibitem{lp}
E.~M.~Lifshitz and L.~P.~Pitaevski, {\it Statistical physics, Part 2}
(Pergamon, New York, 1987).


\bibitem{81}
H.~Boschi-Filho, C.~P.~Natividade and C.~Farina, 
Phys.\ Rev.\ D {\bf 45}, 586 (1992).


\bibitem{Temp}
M.~D'Elia, A.~Di Giacomo and E.~Meggiolaro, 
Phys.\ Rev.\ D {\bf 67}, 114504 (2003)
[arXiv:hep-lat/0205018].


\bibitem{ep}
B.~Svetitsky and L.~G.~Yaffe,
Nucl.\ Phys.\ B {\bf 210}, 423 (1982);
J.~Engels, J.~Jersak, K.~Kanaya, E.~Laermann, C.~B.~Lang, T.~Neuhaus and H.~Satz,
Nucl.\ Phys.\ B {\bf 280}, 577 (1987);
A.~Di Giacomo, M.~Maggiore and S.~Olejnik,
Nucl.\ Phys.\ B {\bf 347}, 441 (1990).


\bibitem{chir}
Yu.~A.~Simonov,
Sov.\ J.\ Nucl.\ Phys.\  {\bf 50}, 134 (1989)
[Yad.\ Fiz.\  {\bf 50}, 213 (1989)].

\bibitem{L}
T.~Doi, N.~Ishii, M.~Oka and H.~Suganuma,
Phys.\ Rev.\ D {\bf 70}, 034510 (2004)
[arXiv:hep-lat/0402005].


\bibitem{chir1}
J.~Babington, J.~Erdmenger, N.~J.~Evans, Z.~Guralnik and I.~Kirsch,
Phys.\ Rev.\ D {\bf 69}, 066007 (2004)
[arXiv:hep-th/0306018].


\end{thebibliography}
\end{document}